\documentclass[12pt,a4paper]{article}
\usepackage{graphicx}
\usepackage{times}
\title{\bf Recent ProAm Campaigns: \\Be stars, {\it COROT} and others}
\author{Jose Ribeiro\\
\\
\normalsize jmscrib@gmail.com - ConVento Group
\\
\\
\normalsize Published in proceedings of \\
\normalsize"Stellar Winds in Interaction", editors T. Eversberg and J.H. Knapen. \\ 
\normalsize Full proceedings volume is available on http://www.stsci.de/pdf/arrabida.pdf
}
\date{\mbox{}}
\begin{document}
\maketitle
%
%
\def\bull{\vrule height .9ex width .8ex depth -.1ex}
\makeatletter
\def\ps@plain{\let\@mkboth\gobbletwo
\def\@oddhead{}\def\@oddfoot{\hfil\tiny\bull\quad
Workshop ``Stellar Winds in Interaction'' Convento da Arr\'abida, 2010 May 29 - June 2\quad\bull}%
\def\@evenhead{}\let\@evenfoot\@oddfoot}
\makeatother
%
%
\def\beginrefer{\section*{References}%
\begin{quotation}\mbox{}\par}
\def\refer#1\par{{\setlength{\parindent}{-\leftmargin}\indent#1\par}}
\def\endrefer{\end{quotation}}
%
%
%
%
\section{Introduction}
The ProAm effort in modern astronomy and astrophysics is now a reality. The achievements of amateur astronomers throughout the early history of astronomy are well known. However, during the 20th century, the high specialisation and the technology required for dealing with the astronomical issues of that epoch, forced a natural separation of the professional activities from the amateur ones. Nowadays, technology has become much cheaper. Amateur astronomers have equipment capable of producing scientific results within their reach. Good equipment pushes some amateur astronomers to learn more, recovering their dialog with professionals. 
ProAm activities may be fruitful in data mining, in database feeding, in long-term campaigns and even in casual observations.

\section{The French Effort}

The French effort in the development of ProAm collaborations is paradigmatic. In 2003, the CNRS and the amateur association AUDE organised a school of astrophysics for amateurs with the aim of preparing collaborators for some of their projects, namely the ground-based support for a mission of {\it COROT} on the physics of Be stars'. This school resulted in the development of an affordable spectrograph, able to do science. At the end of 2005, 76 spectrograph kits of the LHIRESIII type were distributed. LHIRESIII
is of a Littrow design and achieves a resolution of 17000 in H$\alpha$. In May 2006, a new school was organised in order to teach the use of the spectrograph, as well as all the reduction and calibration procedures. At this school, two long-term collaborations were launched, one on RR Lyrae, and another on Be stars. A database on Be stars was in preparation. In December 2006, {\it COROT} was launched. During the first trimester of 2007 the database of Be Star Spectra (BeSS, http://basebe.obspm.fr/basebe/), led by the Paris-Meudon Observatory, was available on the web for both professionals and amateurs who wished to deliver Be star spectra. In August 2007, at a practical spectroscopy course at the Observatoire de Haute-Provence (OHP), some brainstorming was done in order to organise the Be star observations. Issues as observing frequency, activity alerts, and observer night preparation were discussed. At the OHP course in June 2008 a very useful tool, the Arasbeam, was born (http://arasbeam.free.fr/?lang=en). It is a web site that each observer consults in order to choose the most urgent stars to be observed. This way, the community maximises the coverage of these stars.
Thanks to that, the results speak for themselves:

\begin{itemize}
	\item Eight Be activities were detected since then.
	\item 10640 amateur spectra in BeSS.
	\item 476 different Be stars $m(V) <$ 9, dec $> -$25 (since 6/2008, after ArasBeam implementation).
	\item Only 144 different Be stars were in BeSS before 5/2008!
	\item Only 60 Be stars of $m(V) <$ 9 don't have any spectrum.
	\item All stars of $m(V) <$ 7 have at least one spectrum.
	\item From the {\it COROT} mission on Be stars, two refereed papers resulted with amateur co-authors (Guti\'errez-Soto et al. 2009; Neiner et al. 2009).
	\item The French site ARAS\footnote{http://www.astrosurf.com/aras/} also organised some long-term observational campaigns on several other targets.
	\item Another refereed paper used BeSS data, and includes an amateur as co-author ({\v S}tefl et al. 2009). 
	
\end{itemize}

\section{The Czech ProAm Collaborations}

Some scientists from the Astronomical Institute of the Czech Republic and from the Astronomical Institute of the Charles University in Prague accept collaborations with amateurs in spectroscopy. Here, some long-term observations such as the follow up of upsilon Sagittarii are being performed. As of 2007, the phase curve of the  radial velocity of the blue side of the H$\alpha$ line was not well defined, with many missing points. Today, the behaviour of that line is well established, at least in the present state of the system.
The Pleione system has been followed by the Czech group for nine years, recurring to the Czech Ondrejov and to the Canadian DAO telescopes. The orbital period of the system was already established by other teams. However, some doubts existed about the orbital eccentricity, and it was important to take some RV measurements during the superior conjunction of the system. At that date, it was overcast in Ondrejov and in Canada. It was clear in Lisbon, and four spectra were taken, three of them in three consecutive nights. This casual and independent observation was helpful for the study, resulting in a refereed publication (Nemravov{\'a} et al. 2010).
Many other examples of ProAm collaborations with the Czech scientists can be found in ADS (e.g., papers co-authored by C. Buil).

\section{Conclusions}

Better equipped and skilled amateur astronomers can make valuable scientific contributions. Today, amateurs have at their reach Echelle spectrographs. By applying algorithms that before were only at the reach of professionals, amateurs can obtain RV results unthinkable a year ago! A 28\,cm telescope equipped with an Echelle, in the suburbs of Toulouse, reached 75\,m/s radial velocity!!\footnote{http://www.astrosurf.org/buil/extrasolar/obs.htm}
The recent campaigns for the coverage of the WR~140 periastron and the still ongoing epsilon Aurigae eclipse are paradigmatic examples of the importance of coordinated work. The role of the secular AAVSO in photometry and of ARAS in spectroscopy are good examples of this. In future, an organization coordinating all ProAm activity across all astrophysical techniques and all wavelengths, aiming for long-term campaigns, is needed. Recently the first steps in that direction were set by the creation of the ConVento group during the workshop ``Colliding Winds in Interaction", held in Portugal\footnote{http://astrosurf.com/joseribeiro/e\_arrabida.htm}. Although exclusively dedicated to stellar astrophysics, it is the first ProAm international group covering all the astrophysical wavelengths and all astrophysical techniques (http://www.stsci.de/convento/).

%
%

%
%
\footnotesize
\beginrefer

\refer Guti{\'e}rrez-Soto, J., et al.\ 2009, A\&A, 506, 133

\refer Neiner, C., et al.\ 2009, A\&A, 506, 143 

\refer Nemravov{\'a}, J., et al.\ 2010, A\&A, 516, A80 

\refer {\v S}tefl, S., et al.\ 2009, A\&A, 504, 929 

\endrefer

          \newpage
          \qquad
          
\end{document}